\documentclass[twocolumn]{aastex631}

\usepackage{comment}

\usepackage{xcolor}

\shorttitle{CO Observations of the Type Ia SNR 3C~397}
\shortauthors{Ito et al. (2023)}
\graphicspath{{./}{figures/}}

\begin{document}

\title{CO Observations of the Type Ia Supernova Remnant 3C~397 by the Nobeyama 45-m Radio Telescope: Possible Evidence for the Single-Degenerated Explosion}

\author[0009-0000-4742-5098]{D. Ito}
\affiliation{Department of Physics, Nagoya University, Furo-cho, Chikusa-ku, Nagoya 464-8601, Japan: ito\underline{ }d@u.phys.nagoya-u.ac.jp}
\affiliation{Faculty of Engineering, Gifu University, 1-1 Yanagido, Gifu 501-1193, Japan: sano.hidetoshi.w4@f.gifu-u.ac.jp}

\author[0000-0003-2062-5692]{H. Sano}
\affiliation{Faculty of Engineering, Gifu University, 1-1 Yanagido, Gifu 501-1193, Japan: sano.hidetoshi.w4@f.gifu-u.ac.jp}
\affiliation{Center for Space Research and Utilization Promotion (c-SRUP), Gifu University, 1-1 Yanagido, Gifu 501-1193, Japan}

\author[0000-0003-2930-350X]{K. Nakazawa}
\affiliation{Kobayashi-Maskawa Institute for the Origin of Particles and the Universe, Nagoya University, Nagoya-shi, Aichi, Japan}

\author[0000-0002-9901-233X]{I. Mitsuishi}
\affiliation{Department of Physics, Nagoya University, Furo-cho, Chikusa-ku, Nagoya 464-8601, Japan: ito\underline{ }d@u.phys.nagoya-u.ac.jp}

\author[0000-0002-8966-9856]{Y. Fukui}
\affiliation{Department of Physics, Nagoya University, Furo-cho, Chikusa-ku, Nagoya 464-8601, Japan: ito\underline{ }d@u.phys.nagoya-u.ac.jp}

\author{H. Sudou}
\affiliation{
National Institute of Technology, Sendai College 48 Nodayama, Medeshima-Shiote, Natori, Miyagi 981-1239, Japan}
\affiliation{Faculty of Engineering, Gifu University, 1-1 Yanagido, Gifu 501-1193, Japan: sano.hidetoshi.w4@f.gifu-u.ac.jp}

\author{H. Takaba}
\affiliation{Faculty of Engineering, Gifu University, 1-1 Yanagido, Gifu 501-1193, Japan: sano.hidetoshi.w4@f.gifu-u.ac.jp}



\begin{abstract}

We present a new CO observation toward the Type Ia supernova remnant (SNR) 3C~397 using the Nobeyama 45-m radio telescope at an unprecedent angular resolution of $\sim$18$\arcsec$. We newly found that the CO cloud at $V_{\mathrm{LSR}}$ = 55.7--62.2~km~s$^{-1}$ (60~km~s$^{-1}$ cloud) shows a good spatial correspondence with the radio continuum shell. We also found an expanding gas motion of the 60~km~s$^{-1}$ cloud with an expansion velocity of $\sim$3~km~s$^{-1}$, which is thought to be formed by the pre-and/or post-supernova feedback. By considering the positions of Galactic spiral arms and the X-ray/H{\sc i} absorption studies, we concluded that 3C~397 is physically associated with the 60~km~s$^{-1}$ cloud rather than the previously known CO cloud at $V_{\mathrm{LSR}}$ $\sim$30~km~s$^{-1}$. Given that the previously measured pre-shock density is $\sim$2--5~cm$^{-3}$, the expanding motion of the 60~km~s$^{-1}$ cloud was likely formed by the pre-supernova feedback known as optically thick wind. The scenario is consistent with that 3C~397 exploded inside a wind-blown bubble as a single degenerate system.

\end{abstract}



\section{INTRODUCTION}\label{sec:intro}

Identifying progenitor systems of Type Ia supernovae (SNe) is a pressing issue in modern astrophysics. This is because it is the standard candle in the universe due to its constant absolute luminosity \cite[e.g.,][]{1977SvA....21..675P, 1993ApJ...413L.105P, 1999AJ....118.1766P}. ``Single-Degenerate (SD)'' and ``Double-Degenerate (DD)'' are the two most widely accepted scenarios to describe the progenitor systems of Type Ia SNe. The SD scenario is the accretion of gas from a companion star onto a white dwarf, reaching the Chandrasekhar limit, and an ensues thermonuclear explosion \cite[e.g.,][]{1973ApJ...186.1007W, 1982ApJ...253..798N, 1984ApJ...286..644N}. On the other hand, the DD scenario is an explosion caused by merging two white dwarfs \cite[e.g.,][]{1984ApJ...277..355W, 1984ApJ...284..719I}. Only for the SD scenario, there exists a companion star blown away by the explosion, so it is the simplest evidence to identify the companion star. However, there is no strong case of observational identification of the companion star, although many studies have been performed \cite[e.g.,][]{2014ARA&A..52..107M, 2016IJMPD..2530024M}.

An alternative way to verify the SD scenario is to find an expanding shell interacting with a Type Ia supernova remnant (SNR). Because the expanding gaseous shell surrounding a Type Ia SNR could be formed by disk winds, referred to as ``optically thick wind (OTW)'', from the progenitor of a binary system comprising a white dwarf and a nondegenerate companion \cite[e.g.,][]{1996ApJ...470L..97H, 1999ApJ...522..487H, 2000ApJ...528L..97H}. When the mass accretion rate of the ISM from the companion to the white dwarf exceeds a critical value, strong winds blow to stabilize mass transfer, forming a low-density bubble, and then the supernova explosion occurs in it \citep{2007ApJ...662..472B}. The shell size formed by the OTW depends not only on the accretion wind parameters but also on the gas density of the surrounding medium \citep{1992ApJ...388...93K, 1992ApJ...388..103K}. A series of processes are observed only in the SD scenario, but not in the DD scenario.

Such attempts to explain SD scenarios in terms of an expanding shell have been performed in previous studies. The first detection of an expanding shell was the $^{12}$CO observation of SN~1572 \citep{2016ApJ...826...34Z}. They found an expanding shell with mass $\sim$220~$M_{\sun}$ and expansion velocity $\Delta V\sim$5~km~s$^{-1}$ and argued that its momentum is sufficiently explained by energy injection from the OTW. They concluded that SN~1572 was produced by the SD scenario. Other Type Ia SNRs such as N103B, G344.7$-$0.1, and SN~1006 also showed expanding shells possibly due to the OTW \citep{2018ApJ...867....7S, 2022ApJ...933..157S, 2020ApJ...897...62F}. To explain the progenitor systems of Type Ia SNe in the SD scenario, further CO observations in other Type Ia SNRs are needed to test the scenario.

3C~397 (G41.1$-$0.3) is a bright Type Ia SNR in X-ray and radio wavelengths at ($l$, $b$)$\sim$(41$\fdg$1, $-$0$\fdg$3). The SNR is classified as a mixed-morphology type whose central region is bright in thermal X-rays and distributed on a shell in radio continuum emission, which shows a rectangular shell elongated in the east-west direction with an apparent size of 4$\farcm$5 $\times$ 2$\farcm$5 \cite[e.g.,][]{1985ApJ...296..461B, 1993ApJ...408..514A, 1998ApJ...503L.167R, 2005ApJ...618..321S}. The age discrepancies are $\sim$1350--1500~yr for radio \citep{2016ApJ...817...74L} and $\sim$5000~yr for X-ray observations \citep{2005ApJ...618..321S}.

3C~397 has been observed in radio, infrared, X-ray, and gamma rays \cite[e.g.,][]{2000ApJ...545..922S, 2010ApJ...712.1147J, 2015ApJ...801L..31Y, 2021ApJ...913L..34O, 2019AJ....157..123L, 2021MNRAS.501.4226E}. \citet{2019AJ....157..123L} concluded that the SNR morphology is not due to an asymmetric explosion but to an explosion within an inhomogeneous ISM based on near-infrared [Fe{\sc ii}] emission and radio continuum observations. \citet{2015ApJ...801L..31Y} detected K-shell emission lines from Cr, Mn, Fe, and Ni using Suzaku that cannot be explained without electron capture during explosions and found that their mass ratios of Mn/Fe and Ni/Fe are very high, suggesting that 3C~397 has a progenitor of the SD scenario.

3C~397 is thought to be associated with the dense molecular clouds by radio-line observations. \citet{2005ApJ...618..321S} carried out CO($J$~=~2--1) and CO($J$~=~1--0) observations with the Swedish-ESO Submillimeter Telescope (SEST) and the Mopra telescope, whose angular resolutions of $\sim$22$\arcsec$ and $\sim$45$\arcsec$, respectively. The authors found that the molecular clouds at a velocity of $\sim$38--40~km~s$^{-1}$ are well spatially correlated with the X-ray brightness on the western shell of the SNR. Subsequent $^{12}$CO($J$~=~1--0) observations using the 13.7~m telescope at the Purple Mountain Observatory suggested the cloud association at the velocity range of $\sim$27--35~km~s$^{-1}$, except for the southeastern shell of the SNR \citep{2010ApJ...712.1147J}. The authors also found possible evidence for a line-broadening of CO at a velocity of $\sim$32~km~s$^{-1}$\footnote{This line-broadening is also mentioned by \cite{2016ApJ...816....1K} using the Heinrich Submillimeter Telescope (SMT).}. Both the authors estimated the kinematic distance of the clouds to be $\sim$10~kpc, which is consistent with a large absorbing column density of X-rays. On the other hand, there are no detailed CO studies that focus on the other velocity components and kinematics of the gas, e.g., an expanding motion of the clouds. Moreover, the 1420~MHz H{\sc i} absorption studies suggested that the systemic velocity of 3C~397 is $\sim$50--60~km~s$^{-1}$, corresponding to the lower and upper limits of kinematic distances to be 6.3$\pm$0.1 and 9.7$\pm$0.3~kpc, respectively \citep{2016ApJ...817...74L}. If this is correct, the shock-interacting molecular clouds with 3C~397 would need to be reconsidered.

In the present paper, we report new millimeter-wavelength observations using $^{12}$CO($J$~=~1--0) line emission with the Nobeyama 45~m radio telescope (NRO45) and archival $^{12}$CO($J$~=~3--2) data using the James Clerk Maxwell Telescope (JCMT). The CO datasets with unprecedented sensitivity and high angular resolution of $\sim$20$\arcsec$ enable us to identify interacting molecular clouds and their physical relation to the high-energy phenomena in 3C~397. Section~\ref{sec:obs} describes the observational datasets and their reductions. Section \ref{sec:res} comprises five subsections: Sections \ref{subsec:xandradio} and \ref{subsec:distribution_of_co} present overviews of the distributions of X-rays, the radio continuum, and CO; Section \ref{subsec:HR} shows the X-ray hardness ratio. A discussion and conclusions are given in Sections \ref{sec:dis} and \ref{sec:con}, respectively.

\section{OBSERVATIONS AND DATA REDUCTION} \label{sec:obs}
\subsection{CO} \label{subsec:CO}

We performed $^{12}$CO($J$~=~1--0) observations at 115.271202~GHz with NRO45 on December 7, 2022 (proposal No.G22030, PI: H. Sano). We used the on-the-fly mapping mode for a 12$\arcmin$ $\times$ 12$\arcmin$ rectangular region centered at ($l$, $b$) = (41\fdg12, $-0$\fdg31). The front end was the FOur-beam REceiver System on a 45-m Telescope (FOREST: \citealp{10.1117/12.2232137}), a dual-polarization, double sideband, four-beam heterodyne receiver with a superconductor-insulator-superconductor (SIS) mixer driven at 80--116~GHz. The back end was an FX-type digital spectrograph named Spectral Analysis Machine for 45m Telescope (SAM45: \citealp{6051296}). The bandwidth was 125~MHz, and the channel width was 61.04~kHz in the spectral window mode. The typical system temperature was $\sim$340~K for both V and H polarization including the atmosphere. The final beam size was $\sim$18$\arcmin$ in the full-width half-maximum (FWHM). The pointing accuracy was achieved to be better than $\sim2\arcmin$ through hourly observations of the SiO maser OH39.7$+$1.5. We also observed W51[($\alpha_{\mathrm{B}1950}$, $\delta_{\mathrm{B}1950}$) = (19$^{\mathrm{h}}$21$^{\mathrm{m}}$26\fs2, 14\arcdeg24\arcmin43\farcs0)] for the absolute intensity calibration, and then we estimated the extended main beam efficiency of $\sim$0.56 through a comparison with the archival FUGIN data \citep{2017PASJ...69...78U}. The final data has a noise fluctuation of $\sim$0.11~K at the velocity resolution of 0.65~km~s$^{-1}$.

We also used archival data of $^{12}$CO($J$~=~3--2) line emission at 345.795990~GHz observed with JCMT by the $^{12}$CO($J$~=~3--2) High-Resolution Survey (COHRS DR2: \citealp{2023ApJS..264...16P}). The angular resolution is $\sim16\farcs6$ in the FWHM. We applied the main beam efficiency of $\sim$0.61 to convert the data from antenna temperature $T_{\mathrm{a}}^{\ast}$ to main-beam temperature $T_{\mathrm{mb}}$\citep{2023ApJS..264...16P}. The final data has a noise fluctuation of $\sim$0.11~K at the velocity resolution of 0.65~km~s$^{-1}$.

\subsection{Radio Continuum} \label{subsec:RC}
To compare CO distributions with the SNR shell morphology, we used the 20~cm radio continuum data taken with the Very Large Array (VLA) as a part of the Multi-Array Galactic Plane Imaging Survey (MAGPIS: \citealp{2006AJ....131.2525H}). The angular resolution is 6$\farcs$4 $\times$ 5$\farcs$4 with a position angle of 350$\arcdeg$, and the typical noise fluctuation is $\sim$0.3 mJy beam$^{-1}$.

\subsection{X-rays} \label{subsec:X}
We used archival X-ray data obtained by Chandra with the ACIS-I array (Obs ID:~1042), which has been published by several papers \citep[e.g.,][]{2005ApJ...618..321S, 2010ApJ...712.1147J, 2016ApJ...821...20K}. We used CIAO version 4.14 \citep{10.1117/12.671760} with CALDB 4.9.4 \citep{10.1117/12.672876} for data reduction and imaging. After reprocessing all data using the {\texttt{chandra\_repro}} procedure, we created energy-filtered, exposure-corrected maps using the {\texttt{flux\_image}} procedure in the energy bands of 0.5--7.0~keV (broadband), 1.14--1.27~keV (soft-band), and 3.3--3.7~keV (hard-band). The effective exposure time was $\sim$66~ks. We smoothed each image to improve the signal-to-noise ratio: the final beam size of the soft and medium band images is to be $\sim$10$\arcsec$ and that of the broadband image is $\sim$3$\arcsec$ in the FWHM.

\begin{figure}
\includegraphics[width=\linewidth,clip]{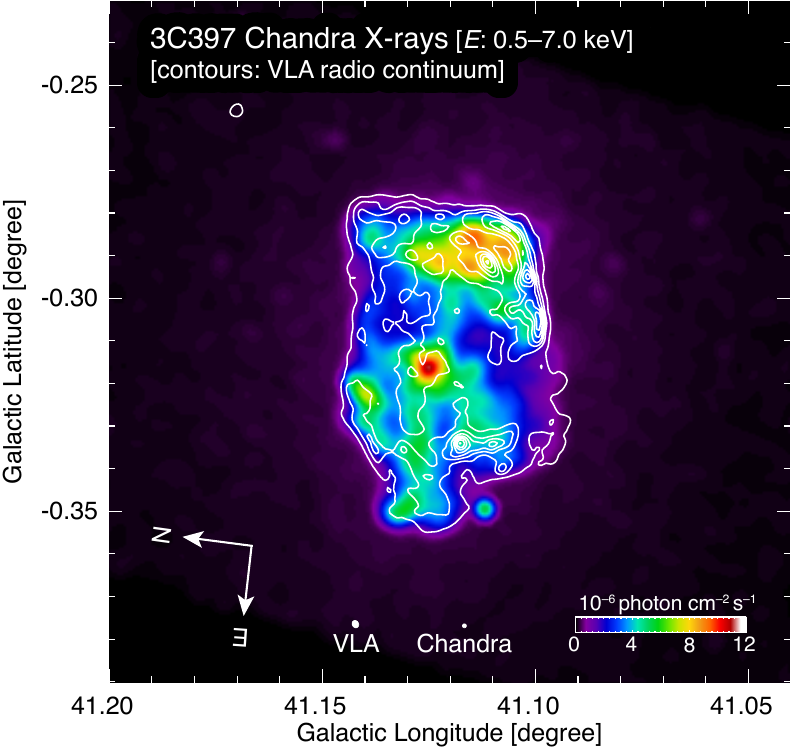}
\caption{Distribution of Chandra X-ray intensity ($E$:~0.5--7.0~keV; \citealp[e.g.,][]{2005ApJ...618..321S}) superposed on the VLA radio continuum at 1.4~GHz \citep{2006AJ....131.2525H}. The lowest contour level and intervals are 1.5 and 7.5~mJy~beam$^{-1}$, respectively.
\label{fig:Xray}}
\end{figure}

\section{RESULTS} \label{sec:res}
\subsection{Overview of X-ray and Radio Continuum} \label{subsec:xandradio}
Figure~\ref{fig:Xray} shows the Chandra X-ray image at 0.5--0.7~keV superposed on the VLA radio continuum contours at 1.4~GHz. As presented in previous studies, X-rays are concentrated in the center of the SNR and show the elongated distribution in the east-west direction \citep[e.g.,][]{1998ApJ...503L.167R}. The radio continuum shows enhancement in the southeast and southwest, consistent with an X-ray distribution brighter than $\sim$5$\times$ 10$^{-6}$~photons~cm$^{-2}$~s$^{-1}$. Similarly, in the northeastern part of the SNR, there is an area of slightly higher intensity in X-rays ($\sim$7$\times$10$^{-6}$~photons~cm$^{-2}$~s$^{-1}$), corresponding to the distribution of the radio shell. No significant enhancement of radio emission was observed in the northwest direction. The intensity distributions of both the radio and X-rays extend east to west from the center of the SNR, which were mentioned as ``jet-like structures'' by \citet{2005ApJ...618..321S}.

\begin{figure*}
\includegraphics[width=\linewidth,clip]{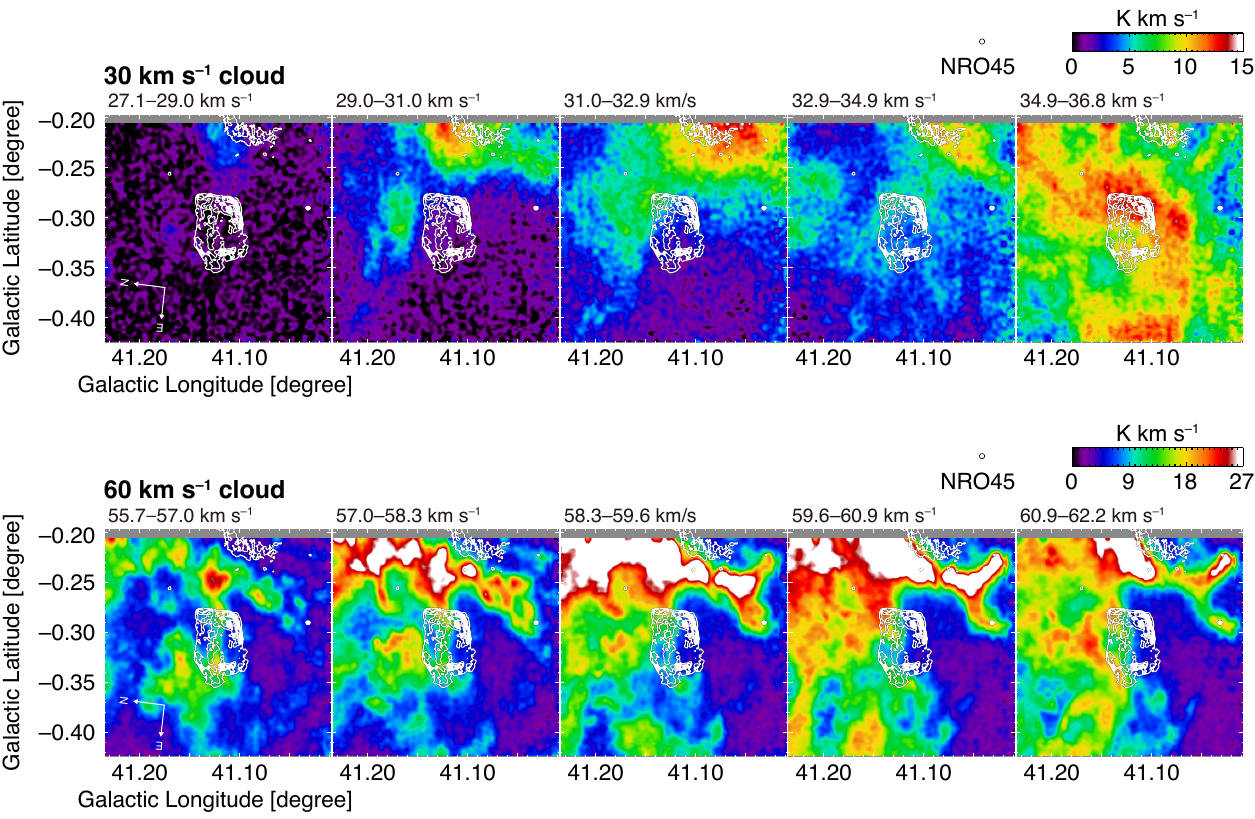}
\caption{Velocity channel distributions of the NRO45 $^{12}$CO($J$~=~1--0) for the 30~km~s$^{-1}$ cloud (upper panels) and the 60~km~s$^{-1}$ cloud (lower panels).
Each panel shows the CO integrated intensity distribution integrated over the velocity range of 27.1--36.8~km~s$^{-1}$ every 1.95~km~s$^{-1}$ for the 30~km~s$^{-1}$ cloud, 55.7--62.2~km~s$^{-1}$ every 1.3~km~s$^{-1}$ for the 60~km~s$^{-1}$ cloud. The superposed contours are the same as those shown in Figure \ref{fig:Xray}.
\label{fig:channnel}}
\end{figure*}

\subsection{Distributions of CO clouds} \label{subsec:distribution_of_co}
Figure \ref{fig:channnel} shows the velocity channel maps of $^{12}$CO($J$~=~1--0) toward 3C~397 for two velocity ranges of 27.1--36.8~km~s$^{-1}$ (hereafter the ``30~km~s$^{-1}$ cloud'') and 55.7--62.2~km~s$^{-1}$ (hereafter the ``60~km~s$^{-1}$ cloud''). The 30~km~s$^{-1}$ cloud is believed to be associated with the SNR \citep{2005ApJ...618..321S, 2010ApJ...712.1147J}, while the 60~km~s$^{-1}$ cloud represents the systemic velocity of 3C~397 by the H{\sc i} absorption study \citep{2016ApJ...817...74L}.

In the 30~km~s$^{-1}$ cloud (Figure \ref{fig:channnel} upper panels), $^{12}$CO emission partially surrounds the northern and western boundaries of the radio shell in the velocity range of 29.00--32.9~km~s$^{-1}$. On the other hand, the $^{12}$CO cloud at the velocity range of 34.9--36.8~km~s$^{-1}$ is almost filled inside the radio shell. In the 60~km~s$^{-1}$ cloud (Figure \ref{fig:channnel} lower panels), we also found that clumpy and diffuse molecular clouds are distributed along the northern half of the radio shell. We also noted that a filamentary CO cloud elongated east and west from the center of the SNR ($V_{\mathrm{LSR}}$: 55.7--58.3~km~s$^{-1}$) seems to overlap with the jet-like structure of the radio continuum and X-rays (see Figure \ref{fig:Xray}). 

\begin{figure*}
\includegraphics[width=\linewidth,clip]{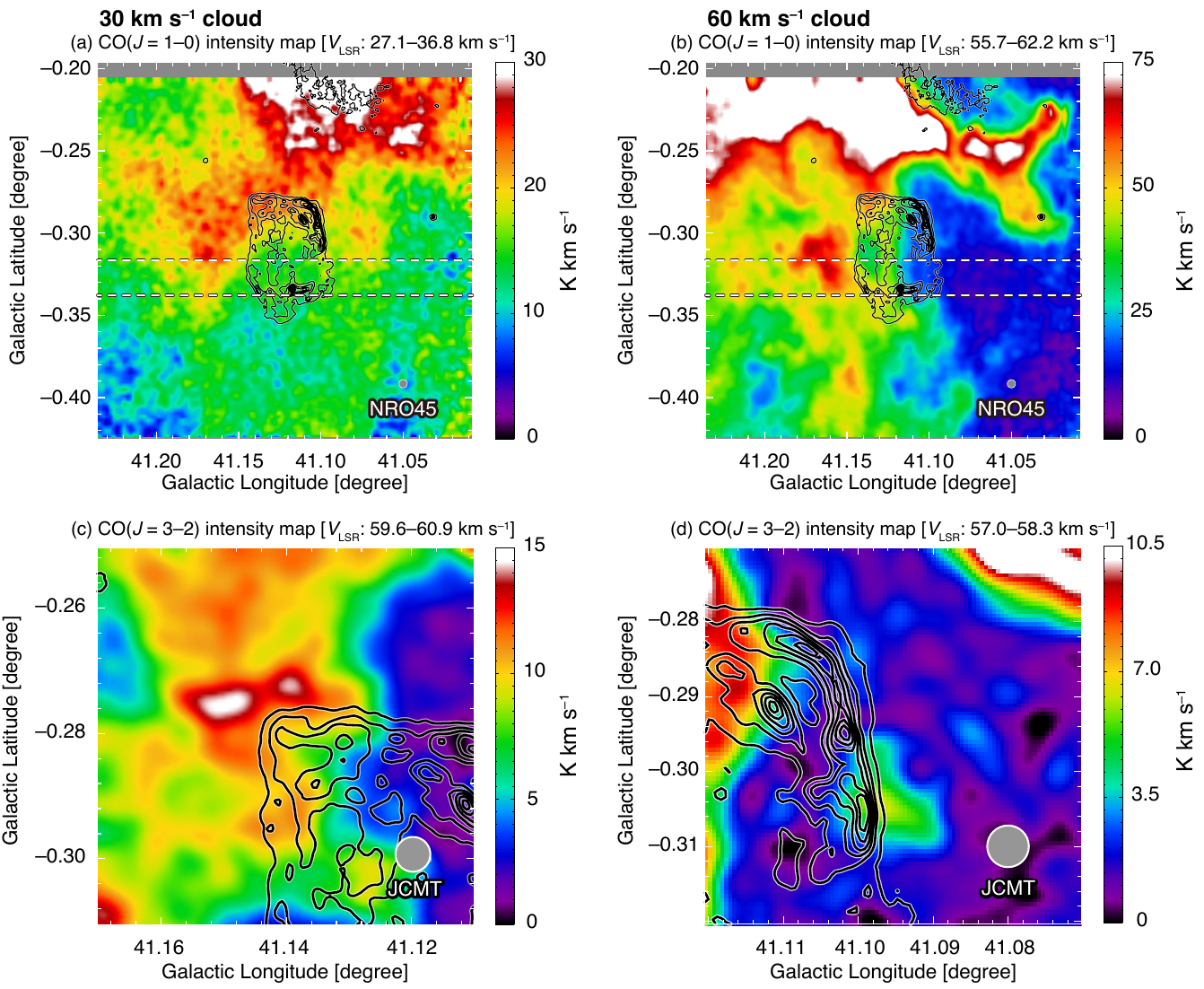}
\caption{(a, b) Integrated intensity maps of NRO45 $^{12}$CO($J$~=~1--0) for (a) the the 30~km~s$^{-1}$ cloud and (b) the 60~km~s$^{-1}$ cloud. The integration velocity range of 27.1--36.8~km~s$^{-1}$ for the 30~km~s$^{-1}$ cloud and 55.7--62.2~km~s$^{-1}$ for the 60~km~s$^{-1}$ cloud. Horizontal dash lines represent the integration ranges of Galactic latitude for each cloud (see Figure \ref{fig:pv}). (c, d) Enlarged maps of the northwestern and southern shells of the 60~km~s$^{-1}$ cloud, which were created by using the archival JCMT $^{12}$CO($J$~=~3--2) data \citep{2023ApJS..264...16P}. The superposed contours are the same as those shown in Figure \ref{fig:Xray}. \label{fig:clouds}}
\end{figure*}

Figures \ref{fig:clouds}(a) and \ref{fig:clouds}(b) show the velocity-integrated intensity maps of $^{12}$CO($J$~=~1--0) for the 30 and 60~km~s$^{-1}$ clouds, respectively. The 30~km~s$^{-1}$ cloud extends the western half of the SNR shell, while the 60~km~s$^{-1}$ cloud is mainly lies on the northern half of the SNR. We also find clumpy $^{12}$CO($J$~=~3--2) clouds in the 60~km~s$^{-1}$ cloud, some of which are along with the radio shell boundary. 

Figures \ref{fig:clouds}(c) and \ref{fig:clouds}(d) show enlarged maps of the northwestern and southwestern parts of the SNR using the JCMT $^{12}$CO($J$~=~3--2) line data. The velocity range is 59.6--60.0 km s$^{-1}$ for Figure \ref{fig:clouds}(c) and 57.0--58.3 km s$^{-1}$ for Figure \ref{fig:clouds}(d). Both the molecular clouds are nicely along the radio-bright edge or shell. Particularly, the dogleg shape of the northwestern radio shell is perfectly along the clumpy cloud distribution.

\begin{figure*}
\includegraphics[width=\linewidth,clip]{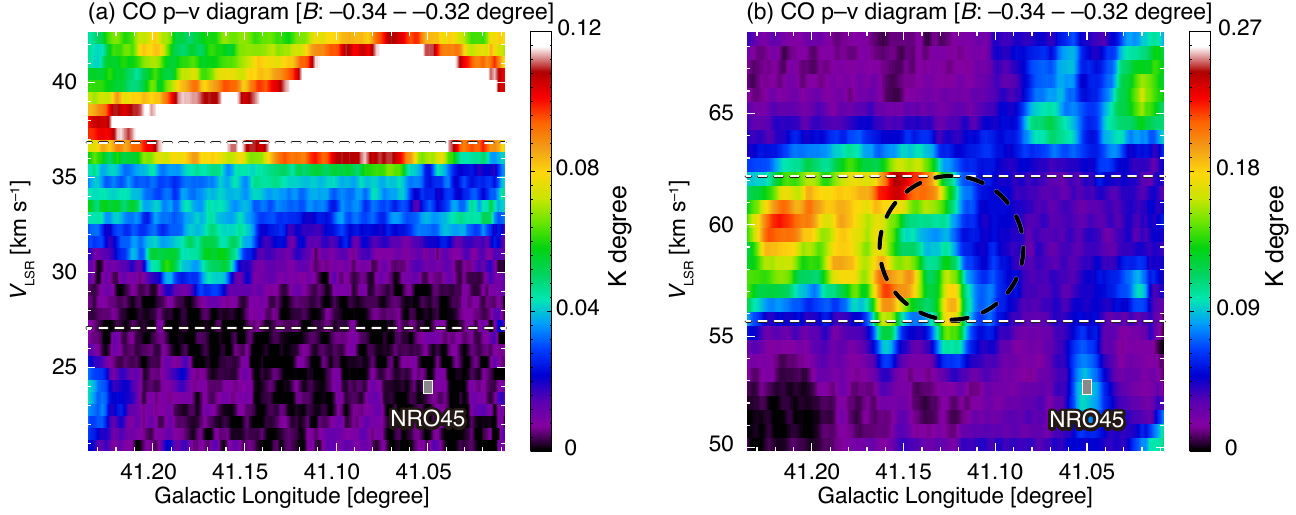}
\caption{Position--velocity diagrams of NRO45 $^{12}$CO($J$~=~1--0) for (a) the 30~km~s$^{-1}$ cloud and (b) the 60~km~s$^{-1}$ cloud. The integration range of Galactic longitude is from $-$0\fdg34 to $-$0\fdg32. The dashed circle in (b) indicates an expanding gas motion. Horizontal dash lines represent the integration velocity ranges for each cloud.
\label{fig:pv}}
\end{figure*}

Figures \ref{fig:pv}(a) and \ref{fig:pv}(b) show the position--velocity for the two molecular clouds. The 30~km~s$^{-1}$ cloud shows no apparent distribution which corresponds to the spatial extent of the SNR. On the other hand, the 60~km~s$^{-1}$ cloud shows a cavity-like distribution in the p--v diagram of $^{12}$CO emission, whose velocity range is from 56 to 62~km~s$^{-1}$. It is noteworthy that the spatial extent of the cavity is roughly consistent with the diameter of the SNR.

\begin{figure*}
\begin{center}
\includegraphics[width=\linewidth,clip]{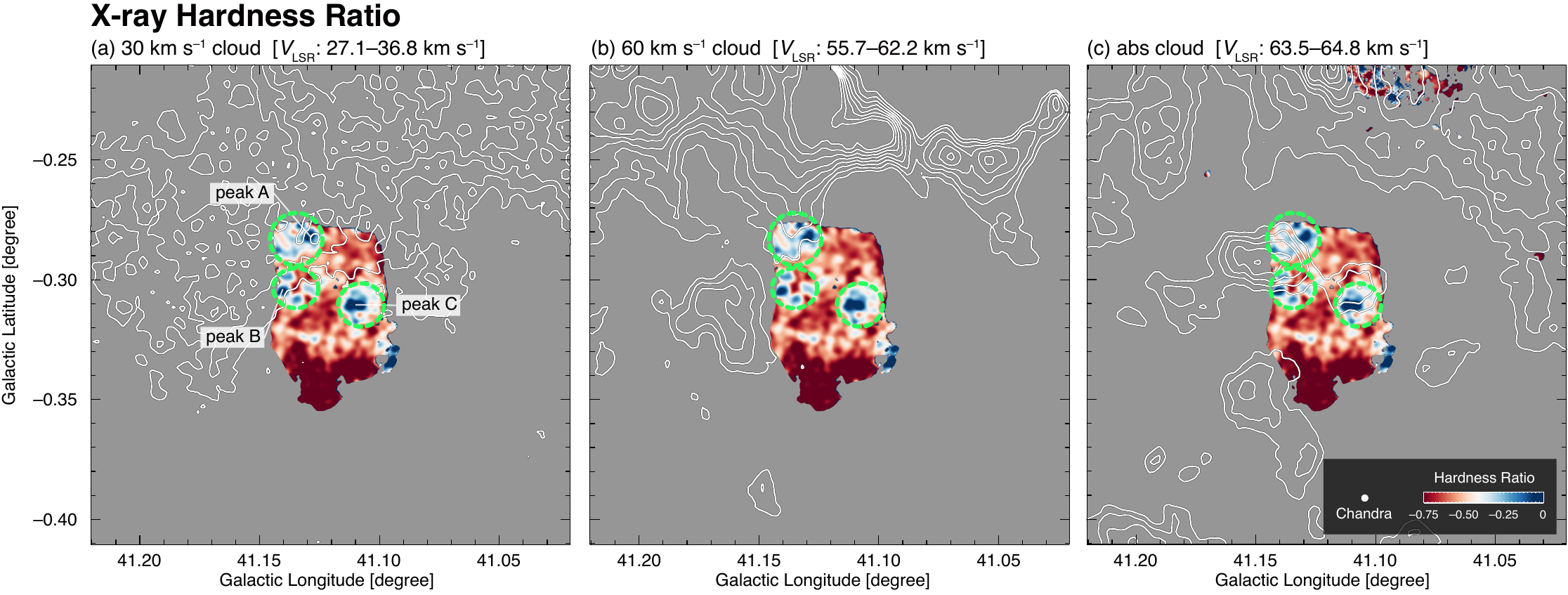}
\end{center}
\caption{X-ray hardness ratio maps overlaid with the NRO45 $^{12}$CO($J$~=~1--0) contours of (a) the 30~km~s$^{-1}$ cloud, (b) the 60~km~s$^{-1}$ cloud, and (c) the abs cloud. The energy band of the hardness ratio map is 1.14--1.27~keV for the soft-band and 3.3--3.7~keV for the hard-band. The shaded areas are excluded because of outside the radio continuum shell. The white contours indicate the $^{12}$CO($J$~=~1--0) integrated intensities, whose lowest contour level and contour intervals are 14.0 and 2.8 K~km~s$^{-1}$ for the 30 km~s$^{-1}$ cloud; 50.0 and 6.0 K~km~s$^{-1}$ for the 60 km~s$^{-1}$ cloud; and 7.5 and 2.0 K~km~s$^{-1}$ for the abs cloud, respectively. The green dashed circles, named peaks A, B, and C, represent the three regions with relatively higher hardness ratio.}
\label{fig:hr}
\end{figure*}

\subsection{Comparison with an X-ray Hardness Ratio} \label{subsec:HR}
In order to clarify a distance relation between the molecular clouds in the line-of-sight direction of 3C~397, we examined the Hardness Ratio ($HR$) of the X-ray. $HR$ is the ratio of X-ray intensities defined as
\begin{equation}
HR=\frac{F_{\mathrm{high}}-F_{\mathrm{low}}}{F_{\mathrm{high}}+F_{\mathrm{low}}}
\end{equation}
where $F_{\mathrm{high}}$ is the intensity of X-ray in the high energy band and $F_{\mathrm{low}}$ is that of the low energy band. Since the soft-band X-rays should be absorbed by dust grains inside a molecular cloud via photoelectric absorption, we can expect a large $HR$ value when dense molecular clouds are located between the SNR and observers in the line-of-sight direction. So, we searched the dense molecular clouds which are spatially corresponding to the larger value of $HR$.

Figure~\ref{fig:hr} shows the $HR$ images superposed on the $^{12}$CO($J$~=~1--0) contours for three different velocity ranges. We found three regions of relatively high $HR$. These coordinates are ($l$, $b$)$\sim$(41\fdg13, $-$0\fdg28), (41\fdg14, $-$0\fdg31), and (41\fdg11, $-$0\fdg31), defined in the order as peak A, peak B, and peak C. The dense molecular cloud (hereafter referred to as ``abs cloud'' in Figure~\ref{fig:hr}(c)) shows that the abs cloud aligns well with all $HR$ peaks, whereas the 30~km~s$^{-1}$ cloud aligns only with peaks A and B, and the 60~km~s$^{-1}$ cloud with peak B alone. No other molecular clouds with good spatial counterparts other than these in the other velocity ranges are found.

\section{Discussion} \label{sec:dis}
\subsection{Molecular Clouds Associated with 3C~397} \label{subsec:MC}
Previous studies proposed that the 30~km~s$^{-1}$ cloud is physically interacting with the SNR 3C~397 \cite[e.g.,][]{2005ApJ...618..321S, 2010ApJ...712.1147J, 2016ApJ...816....1K}. Their argument is based on three observational signatures: (1) the bow-shaped spatial distribution of $^{12}$CO($J$~=~1--0), (2) the line broadening of $\sim$7~km~s$^{-1}$ in the $^{12}$CO($J$~=~1--0) spectrum, and (3) the density of molecular gas assuming a pressure balance between hot and cold gas. In this section, we discuss which of the two clouds, the 30~km~s$^{-1}$ cloud and the newly found 60~km~s$^{-1}$ cloud, is associated with 3C~397 in terms of spatial/velocity distributions and the positional relation among the molecular clouds and 3C~397 toward the line-of-sight.

First, we emphasize that it is very difficult to clarify which clouds are physically associated with 3C~397 in terms of spatial comparison alone. Both the 30 and 60~km~s$^{-1}$ clouds show good spatial correspondence with the radio continuum shell of 3C~397. The diffuse gas of the 30~km~s$^{-1}$ cloud appears to be along the western half of the shell. The $^{12}$CO($J$~=~3--2) clumpy structure in the 60~km~s$^{-1}$ cloud is located in the vicinity of the non-thermal radio-bright shell, indicating that the shock-cloud interaction with magnetic field amplification occurred \citep[e.g.,][]{2009ApJ...695..825I, 2012ApJ...744...71I, 2010ApJ...724...59S, 2013ApJ...778...59S}.

Next, we argue that the cavity-like structure in the p--v diagram of the 60~km~s$^{-1}$ cloud represents an expanding gas motion due to the pre-and/or post-supernova feedback effect. The pre-supernova feedback corresponds to strong winds from the massive progenitor or OTW from a binary star system including a white dwarf \citep[e.g.,][]{1977ApJ...218..377W, 1996ApJ...470L..97H, 1999ApJ...522..487H, 1999ApJ...519..314H}. The post-supernova feedback represents the gas acceleration and/or distraction of supernova shocks \citep[e.g.,][]{1990ApJ...364..178K, 1998ApJ...505..286S, 2004AJ....127.1098S}. In the case of 3C~397, the expansion velocity of the gas was estimated to be $\Delta V\sim$3~km~s$^{-1}$ centered at the systemic velocity of $\sim$59 $\pm$ 2~km~s$^{-1}$. The derived expansion velocity is roughly consistent with those of the Galactic SNR SN~1006 (e.g., $\Delta V\sim$4~km~s$^{-1}$, \citealp{2022ApJ...933..157S}), G344.7$-$0.1 (e.g., $\Delta V\sim$4.5~km~s$^{-1}$, \citealp{2020ApJ...897...62F}), and G298.6$-$0.0 (e.g., $\Delta V\sim$8~km~s$^{-1}$, \citealp{2023PASJ...75..384Y}). On the other hand, the p--v diagram of the 30~km~s$^{-1}$ cloud does not show such an apparent signature of an expanding gas motion.

\begin{figure}
\includegraphics[width=\linewidth,clip]{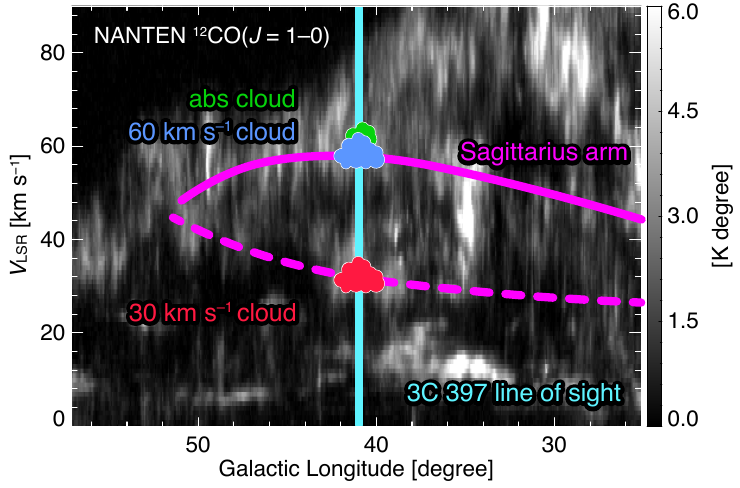}
\caption{Large-scale position--velocity diagram of $^{12}$CO($J$~=~1--0) obtained with the NANTEN 4-m radio telescope \citep{2004ASPC..317...59M}. The integration range in the Galactic Longitude is from $-$1$\arcdeg$ to 1$\arcdeg$. The dashed and solid curves indicate the near and for side Sagittarius arms, respectively. The cyan vertical line represents the line-of-sight direction of 3C~397. We also plotted the 30~km~s$^{-1}$ cloud (red), abs cloud (green), 60~km~s$^{-1}$ cloud (blue).
\label{fig:nanten}}
\end{figure}

\begin{figure}
\includegraphics[width=\linewidth,clip]{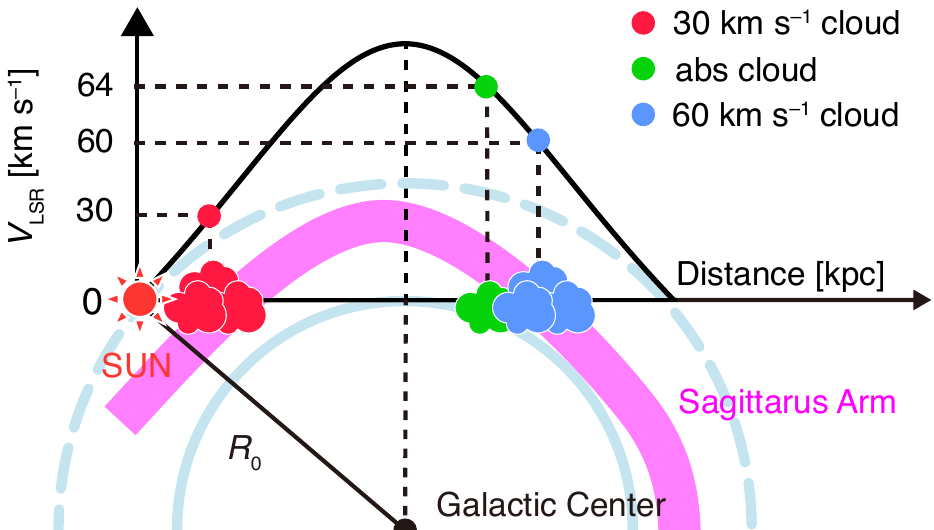}
\caption{Schematic view of the Galactic spiral arm and the distance--velocity relation. The solid black curve shows the relation between the line-of-sight velocity and kinematic distance. The red, green, and blue clouds represent the 30~km~s$^{-1}$ cloud, abs cloud, and 60~km~s$^{-1}$ cloud, respectively. The red thick curve indicates the Sagittarius arm. $R_0$ is the distance from the Galactic center to the sun.
\label{fig:GR}}
\end{figure}

\begin{figure*}
\plotone{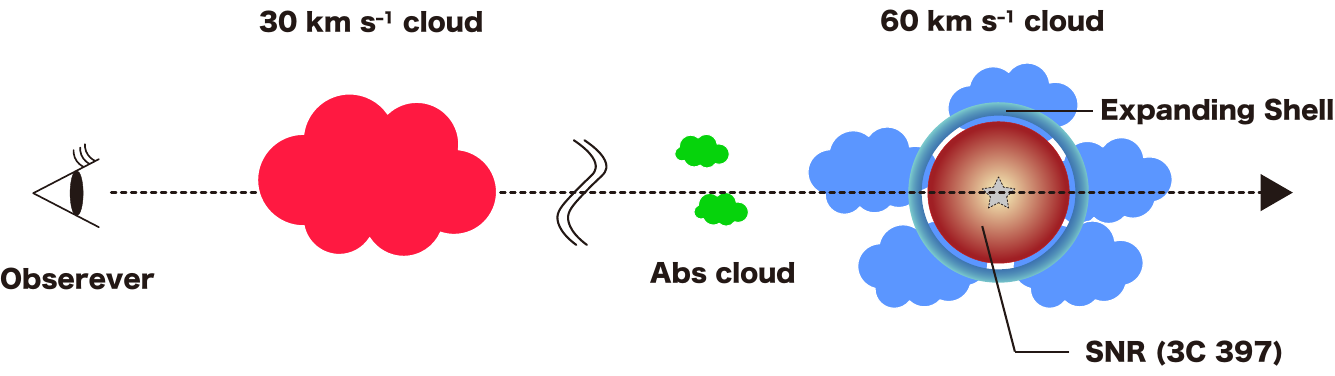}
\caption{Schematic image of the positional relation between each molecular cloud and 3C~397 in the line-of-sight direction. The red, green, and blue clouds represent the 30~km~s$^{-1}$ cloud, abs cloud, and 60~km~s$^{-1}$ cloud, respectively.
\label{fig:LoS}}
\vspace*{1cm}
\end{figure*}
Finally, we discuss the positional relation among the 30~km~s$^{-1}$ cloud, 60~km~s$^{-1}$ cloud, abs cloud ($V_{\mathrm{LSR}}$: 63.5--64.8~km~s$^{-1}$), and 3C~397 toward the line-of-sight direction, in order to check consistency with observational signature, especially in the $HR$ image. We first examined the velocity distribution of the gas in the line-of-sight direction based on the spiral arm distribution. Figure \ref{fig:nanten} shows the large-scale p--v diagram of the $^{12}$CO($J$~=~1--0) obtained using the NANTEN 4-m radio telescope \citep{2004ASPC..317...59M}. According to \citet{2016ApJ...823...77R} the Sagittarius arm, one of the Galactic spiral arms, lies on the line-of-sight direction of 3C~397, whose systemic velocities are $\sim$35~km~s$^{-1}$ (near side arm) and $\sim$60~km~s$^{-1}$ (far-side arm). This implies that the three molecular clouds discussed in this paper are likely on the Sagittarius arm, and their locations are naturally limited to the near-side arm for the 30~km~s$^{-1}$ cloud, and the far-side arm for the 60~km~s$^{-1}$ cloud and abs cloud. Based on the Galactic rotation model by \citet{1993A&A...275...67B}, the kinematic distance is estimated to be $\sim$2.1~kpc for the 30~km~s$^{-1}$ cloud, $\sim$8.3~kpc for the abs cloud, and $\sim$8.7 $\pm$ 0.2~kpc for the 60~km~s$^{-1}$ cloud (see Figure \ref{fig:GR}). From the above discussion, the positional relation among the molecular clouds and 3C~397 in the line-of-sight direction should be expressed as schematically shown in Figure \ref{fig:LoS}. This picture is consistent with the $HR$ study as shown in Section 3.3: the abs cloud is located in front of the SNR because peak C of HR is only spatially correlated with abs cloud. We emphasized that the positional relation is also consistent with the previous H{\sc i} absorption measurement because the absorption line is seen up to the tangent velocity and absence of absorption at $V_{\mathrm{LSR}}$ = 50--60~km~s$^{-1}$ \citep{2016ApJ...817...74L}.

To summarize the above discussion, we conclude that 3C~397 is likely associated with the 60~km~s$^{-1}$ cloud in terms of spatial/velocity distributions of gas (including the previous H{\sc i} absorption study) and the positional relation among the molecular clouds and 3C~397 toward the line-of-sight.


\subsection{Age of 3C~397} \label{subsec:AGE}
The previous section suggested that the distance of 3C~397 is slightly updated from $\sim$10~kpc to $\sim$8.7~kpc. We here estimated the dynamical age $t_{\mathrm{age}}$ of the SNR based on our revised distance of 3C 397. Since the 3C~397 is believed to be in the Sedov--Taylor Phase \citep{1959sdmm.book.....S}, we can estimate the dynamical age as below 
\begin{equation}
 t_{\mathrm{age}} = \frac{2r}{5V_{\mathrm{sh}}},   
\end{equation}
where $V_{\mathrm{sh}}$ is the shock velocity of the SNR and $r$=4.04 pc is the radius of the SNR shell. By adopting to  thw shock velocity $V_{\mathrm{sh}}$ = 1050--1360 km s$^{-1}$ \citep{2016ApJ...817...74L}, we obtain the SNR age $t_{\mathrm{age}}$ = 1100--1400 yr. This value is roughly consistent with the previous age estimation of the 3C~397 that assumed a similar distance \cite[e.g.,][]{2016ApJ...817...74L}.

\subsection{A Hint for a Single-Degenerate Origin} \label{subsec:SD}
In this section, we discuss whether the expanding gaseous shell of the 60~km~s$^{-1}$ cloud was formed before or after the supernova explosion of 3C~397, by comparing the physical properties such as the mass of the evacuated gaseous shell and the pre-shock density. The total mass of the evacuated gaseous shell $M$ could be calculated using the following equations:
\begin{equation}
M=m_{\mathrm{p}}\mu\Omega D^2 \sum_{i}N_{i}(\mathrm{H}_2),
\end{equation}
\begin{equation}
N(\mathrm{H}_2)=X\cdot W(\mathrm{CO}),
\end{equation}
where, $m_{\mathrm{p}}$ is the mass of atomic hydrogen, $\mu$~=~2.8 is the mean molecular weight, $\Omega$ is the solid angle for each data pixel, $D$ is the distance to the SNR, and $N(\mathrm{H}_2)$ is the molecular hydrogen column density, $W$(CO) is the intensity of the $^{12}$CO($J$~=~1--0).

If the derived mass of the expanding shell was uniformly distributed over the present volume of the remnant before being blown out, the pre-shock density of the neutral gas is estimated to be $\sim$800~cm$^{-3}$. On the other hand, \cite{2016ApJ...817...74L} calculated the pre-shock density of $\sim$2--5~cm$^{-3}$ using the previously derived post-shock density from the infrared observation and X-ray emission measure \citep{2005ApJ...618..321S, 2015ApJ...801L..31Y}. 

This discrepancy implies that the expanding gaseous shell in the 60~km~s$^{-1}$ cloud likely had been formed before the supernova explosion of 3C~397. In other words, the supernova explosion of 3C~397 occurred inside a low-density cavity with an expanding velocity of a few~km~s$^{-1}$. In the case of a Type Ia progenitor system like 3C~397, the OTW expected in the SD scenario is the most promising way to form such an expanding shell before the supernova explosion. In fact, Type Ia SNRs N103B and G344.7$-$0.1, which can be understood as the same SD scenario, actually have very similar physical quantities to 3C~397 (e.g., mass and expanding velocity of neutral gas; \citealp{2018ApJ...867....7S, 2020ApJ...897...62F}). Since there are phases in the DD channel that are not yet well understood, this study does not exclude the DD scenario for 3C~397, but it is consistent with the OTW in the SD scenario. A further careful search for a companion star surviving the supernova explosion will shed light on distinguishing the SD and DD scenarios.

\section{Conclusion} \label{sec:con}
We summarize our conclusions as follows:

\begin{enumerate}
\item New $^{12}$CO($J$~=~1--0) observations using the Nobeyama 45~m radio telescope revealed the spatial distributions of two CO clouds at $V_{\mathrm{LSR}}$ = 27.1--36.8~km~s$^{-1}$ (the 30~km~s$^{-1}$ cloud) and 55.7--62.2~km~s$^{-1}$ (the 60~km~s$^{-1}$ cloud) with an unprecedented angular resolution of 18$\arcsec$. We found that both the two clouds show good spatial correspondence with the radio continuum shell of the SNR. The spatial distributions of archival $^{12}$CO($J$~=~3--2) data at the $\sim$20$\arcsec$ resolution also show highly excited cloudlets in the 60~km~s$^{-1}$ cloud, which are nicely along the edge or boundary of the radio-continuum shell.
\item The 60~km~s$^{-1}$ cloud shows a cavity-like structure in a position-velocity diagram, the spatial extent of which is roughly consistent with that of the SNR. This cavity-like structure is generally thought to be an expanding gas motion due to the pre-and/or post-supernova feedback, and hence the 60~km~s$^{-1}$ cloud has a systemic velocity of $\sim$59~km~s$^{-1}$ and an expansion velocity of $\sim$3~km~s$^{-1}$. On the other hand, the 30~km~s$^{-1}$ cloud does not show any such characteristic features.
\item The CO cloud at $V_{\mathrm{LSR}}$ = 62.8--65.4~km~s$^{-1}$ (the abs cloud) lies in the regions where the X-ray hardness ratio shows higher values. This indicates that the abs cloud is located in front of the SNR because the larger value of the X-ray hardness ratio represents the higher photoelectric absorption value toward the line of sight. By considering the distributions of Galactic spiral arms and X-ray hardness ratio, the 30~km~s$^{-1}$ cloud should be located at the near side arm and the others lie on the far-side arm. The kinematic distance cloud be derived using the Galactic rotation curve model to be $\sim$2.1~kpc for the 30~km~s$^{-1}$ cloud, $\sim$8.3~kpc for the abs cloud, and $\sim$8.7~$\pm$ 0.2~kpc for the 60~km~s$^{-1}$ cloud. We,  therefore, concluded that the 60~km~s$^{-1}$ cloud is the one most likely interacting with the SNR.
\item The total gaseous mass within the SNR shell was estimated to be $\sim$$8.6 \times 10^3$~$M_{\sun}$. If all the expanding gas was driven by the supernova shock waves, the pre-shock density was estimated to be $\sim$800~cm$^{-3}$. Since the pre-shock density is inconsistent with the previously estimated value of $\sim$2--5~cm$^{-3}$, 3C~397 likely exploded inside a low-density bubble formed by the optically thick wind. We propose a possible scenario that the progenitor system of 3C~397 is a white dwarf and its companion star, and then a single-degenerate explosion occurred.
\end{enumerate}

We acknowledge Rin Yamada, Yuto Onishi, Hiroki Teramoto, Maki Aruga, Kohei Matsubara, Yurina Yamanaka, and Koharu Hirano for contributions to the observations and data reductions of $^{12}$CO($J$~=~1--0) data. The Nobeyama 45-m radio telescope is operated by the Nobeyama Radio Observatory, a branch of the National Astronomical Observatory of Japan. The James Clerk Maxwell Telescope is operated by the East Asian Observatory on behalf of The National Astronomical Observatory of Japan; Academia Sinica Institute of Astronomy and Astrophysics; the Korea Astronomy and Space Science Institute; the National Astronomical Research Institute of Thailand; Center for Astronomical Mega-Science (as well as the National Key R\&D Program of China with No. 2017YFA0402700). Additional funding support is provided by the Science and Technology Facilities Council of the United Kingdom and participating universities and organizations in the United Kingdom and Canada. The NANTEN project is based on a mutual agreement between Nagoya University and the Carnegie Institution of Washington (CIW). We greatly appreciate the hospitality of all the staff members of the Las Campanas Observatory of CIW. We are thankful to many Japanese public donors and companies who contributed to the realization of the project. This paper employs a list of Chandra datasets, obtained by the Chandra X-ray Observatory, contained in~\dataset[DOI: 10.25574/cdc.187]{https://doi.org/10.25574/cdc.187}. This research has made use of software provided by the Chandra X-Ray Center in the application package CIAO (v4.14). This work was also supported by JSPS KAKENHI grant Nos. 21H01136 and 24H00246 (HS). This work was supported by a University Research Support Grant from the National Astronomical Observatory of Japan (NAOJ). This work was also supported by NAOJ ALMA Scientific Research Grant Code 2023-25A. This research was partially supported by CCI Holdings Co., Ltd. 





\section*{APPENDIX:\\Radius and Thickness of the Expanding Shell}

\begin{figure}[h]
\includegraphics[width=\linewidth,clip]{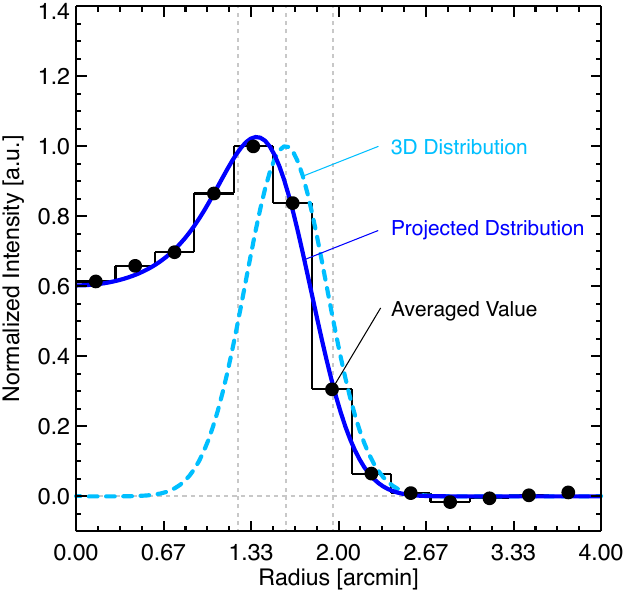}
\caption{Radial profiles of the radio continuum centered at ($l$, $b$) = (41\fdg12, $-$0\fdg31). The black dots and steps are the averaged values of the radio continuum within each annulus. The dashed and solid curves represent the three-dimensional and projected Gaussian distributions, respectively. These curves were obtained by least-squares fitting. The gray dashed vertical lines represent best-fit values for the shell radius and thickness.
\label{gauss}}
\end{figure}

To derive the radius and thickness of the expanding shell, we fitted the radial profile of the radio continuum surface brightness using a 3D spherical shell with a Gaussian function $F(r)$:
\begin{equation}
F(r) = A\exp{\left[-\frac{(r-r_0)^2}{2\sigma^2}\right]}
\end{equation}
where $r$ is the radius of the radio continuum shell and $\sigma$ is the standard deviation of the Gaussian function. We first fitted with a radial profile by moving the origin to determine the geometric center of the SNR. We obtained ($l$, $b$) = (41\fdg03, $-$0\fdg3) as the geometric center with the minimum chi-square value of the least-squares fitting. Figure \ref{gauss} shows the radial profile of the radio continuum at the center position ($l$, $b$) = (41\fdg03, $-$0\fdg3). We obtained $r\sim$4.04~pc (1\farcm60) for the shell radius and $\sigma$ $\sim$1.84~pc (0\farcm73) for the shell thickness as best-fit values, where the shell thickness is defined by the Gaussian FWHM or 2$\sigma\sqrt{2\ln{2}}$. In the present paper, we assumed that the size of the gaseous expanding shell is roughly consistent with that of the radio continuum in the SNR \citep[cf.,][]{2021ApJ...923...15S, 2022ApJ...933..157S, 2022ApJ...938...94A}.


\bibliography{3C397_ref}{}
\bibliographystyle{aasjournal}



\end{document}